\documentclass[12pt]{article}

\pdfoutput=1

\usepackage{amssymb}
\usepackage{amsmath}

\usepackage{color}

\usepackage{graphicx}
\usepackage{amsfonts}
\usepackage{hyperref}
\usepackage{soul}

\newcommand{\vo}{\mathcal{V}}

\def\unit{\relax{\rm 1\kern-.26em I}}
\def\nada{\relax{\rm 0\kern-.30em l}}

\makeatletter
\renewcommand\section{\@startsection {section}{1}{\z@}%
                                 {-3.5ex \@plus -1ex \@minus -.2ex}
                                   {2.3ex \@plus.2ex}%
                                   {\normalfont\large\bfseries}}
\renewcommand\subsection{\@startsection{subsection}{2}{\z@}%
                                   {-3.25ex\@plus -1ex \@minus -.2ex}%
                                     {1.5ex \@plus .2ex}%
                                     {\normalfont\bfseries}}
\renewcommand\subsubsection{\@startsection{subsubsection}{3}{\z@}%
                                   {-3.25ex\@plus -1ex \@minus -.2ex}%
                                     {1.5ex \@plus .2ex}%
                                     {\normalfont\itshape}}
\makeatother

\numberwithin{equation}{section}

\newcommand{\be}{\begin{equation}}
\newcommand{\ee}{\end{equation}}
\newcommand{\bea}{\begin{eqnarray}}
\newcommand{\eea}{\end{eqnarray}}
\newcommand{\barr}{\begin{array}}
\newcommand{\earr}{\end{array}}

\def\beq{\begin{equation}}
\def\eeq{\end{equation}}
\def\be{\begin{equation}}
\def\ee{\end{equation}}
\def\bea{\begin{eqnarray}}
\def\eea{\end{eqnarray}}

\DeclareRobustCommand{\SkipTocEntry}[4]{}

\begin{document}

\begin{titlepage}

\setcounter{page}{1} \baselineskip=19.5pt \thispagestyle{empty}

\begin{flushright}
DESY-13-163\\
\end{flushright}
\vfil

\begin{center}
{\LARGE Low-$\ell$ CMB Power Loss in String Inflation}

\end{center}
\bigskip\

\begin{center}
{\large Francisco G. Pedro and Alexander Westphal}
\end{center}

\begin{center}
\textit{Deutsches Elektronen-Synchrotron DESY, Theory Group, D-22603 Hamburg, Germany}
\end{center} \vfil

\noindent The lack of power on large scales ($\ell\lesssim 40$) might have been observed by the PLANCK satellite. We argue that this putative feature can be explained by a phase of fast roll at the onset of inflation.   We show that in the context of single field models what is required is an asymmetric inflection point model of which fibre inflation is a string motivated example. We study the ability of fibre inflation to generate a suppression of the CMB 2-point function power at low $\ell$, finding that the potential derived from string loops is not steep enough for this purpose. We introduce a steeper contribution to the potential, that dominates away from the inflationary region, and show that if properly tuned it can indeed lead to a spectrum with lack of power at large scales.

\vfil
\begin{flushleft}
September 13, 2013
\end{flushleft}

\end{titlepage}

\newpage
\tableofcontents
\newpage

\section{Introduction}
\label{sec:intro}

Cosmology has long been considered as speculative field. However, several waves of observational progress during the last two decades have required us to fundamentally change that view. At first, highly reliable and increasingly precise measurements of the Hubble Space Telescope (HST) yielded a determination of present-day Hubble parameter $H_0=73.8\pm 2.4 \, \text{km s}^{-1}\text{Mpc}^{-1}$~\cite{Freedman:2000cf,Riess:2011yx}. This milestone enabled the construction of an ultra deep-space distance ladder by type IA supernovae. Systematic observations of such type IA supernovae at high red-shift culminated in the detection of a form of dark energy consistent with extremely small ($\sim 10^{-122}M_{\rm P}^4$) and positive cosmological constant, which drives a late-time accelerated expansion of our Universe~\cite{Riess:1998cb,Perlmutter:1998np}.

In the next step, the space-based satellite missions WMAP~\cite{Hinshaw:2012fq} and PLANCK~\cite{Ade:2013zuv,Ade:2013uln} as well as the ground-based telescopes ACT~\cite{Sievers:2013wk} and SPT~\cite{Story:2012wx,Hou:2012xq} probed the cosmic microwave background (CMB) radiation with unprecedented precision and resolution. Their combined results together with the HST and type IA supernova data led to a concordance model of observational cosmology. In its `essence' this new standard model is consistent with certain simple features of our observed Universe. We find ourselves to a good approximation (sub-\%-level) a FLRW universe which is spatially flat and undergoes accelerated late-time expansion driven by dark energy. Moreover, the recent CMB missions provided measurements of the two-point function power-spectrum  of the $10^{-5}$-level thermal fluctuations with unprecedented precision. These results now bear increasingly strong evidence  that the very early Universe went through a much earlier and extremely rapid phase of accelerated expansion driven by the vacuum energy of a slowly-rolling scalar field, called inflation (see e.g.~\cite{Baumann:2009ds} for a recent review).

The recent high-precision data from PLANCK~\cite{Ade:2013zuv,Ade:2013uln} as well as the ground-based telescopes ACT~\cite{Sievers:2013wk} and SPT~\cite{Story:2012wx,Hou:2012xq}, and in particularly the strong limits on the presence of non-Gaussianity in form of a non-vanishing three-point function from the PLANCK mission~\cite{Ade:2013ydc}, is consistent with a picture of simple slow-roll inflation driven by the scalar potential of a single canonically normalized scalar field. However, the PLANCK results (in continuation of similar features of the WMAP data~\cite{Hinshaw:2012fq}) contain hints at the $2-3\sigma$-level for certain deviations/anomalies at large wavelengths or low $\ell \lesssim 40$ when compared with a simple 'vanilla' universe created from standard slow-roll inflation. One of these possible deviations (and maybe the most persistent one over several different data sets of several CMB probes) consists of lack of CMB two-point function power at low $\ell \lesssim 40$ compared to the concordance model~\cite{Ade:2013uln,Ade:2013nlj}.

One possible explanation for such a lack of power at low-$\ell$ can arise in open inflation~\cite{Linde:1998iw,Linde:1999wv,Yamauchi:2011qq}. Open inflation assumes a pre-inflationary phase consisting of a Coleman-de Luccia (CdL) tunneling transition~\cite{Coleman:1977py,Coleman:1980aw} with a possible subsequent phase of fast-rolling evolution of the inflaton scalar field prior to entering the slow-roll phase. 

In this work we consider the simplest case capable of suppressing curvature perturbation power at large scales -- a fast-roll phase of the inflaton scalar field prior to entering slow-roll~\cite{Linde:1999wv,Yamauchi:2011qq}. The curvature perturbation generated during a phase of quasi-de Sitter expansion behaves as
\beq
\Delta_{\cal R}^2\sim \frac{H^4}{\dot\phi^2}\sim \frac{V}{\epsilon}
\eeq
in slow-roll, and at CMB scales we have typically $\epsilon_{60} \lesssim {\cal O}(0.01)$ generated by slow-roll in a flat region of the scalar potential with a value $V_{60}$. 

Hence, we can generate a loss of power at large scales by having a phase of moderately faster rolling with $\epsilon_{60}<\epsilon_{fast}\lesssim 1$ where still $V\simeq V_{60}$.  A detectable suppression of power at low-$\ell$ can arise in such a situation only if the total amount of e-folds of slow-roll inflation generated after the end of the fast-roll phase is limited to within $2-3$ e-folds of the standard  $N_e^{\rm obs.}\sim 60$ e-folds
\beq\label{NeCond}
N_e^{\rm tot.}\lesssim N_e^{\rm obs.}+{\cal O}(1)\quad.
\eeq
Otherwise the epoch of suppressed curvature perturbation generation redshifts far outside our presently visible horizon by having more then minimum 60 e-folds.

In summary, to achieve low-$\ell$ power suppression we need a scalar potential which just beyond the observable 60-efold field range changes to growing $\epsilon$ very quickly, within a field range corresponding to about $2-3$ e-folds, while keeping $V$ still almost unchanged in the inflationary region.\footnote{For alternative explanations of this phenomenon based on non-standard dynamics of the inflaton at the onset of inflation see \cite{Dudas:2012vv,Sagnotti:2013ica}.}

%
\begin{figure}[t!]
 \centerline{\includegraphics[width=0.7\textwidth]{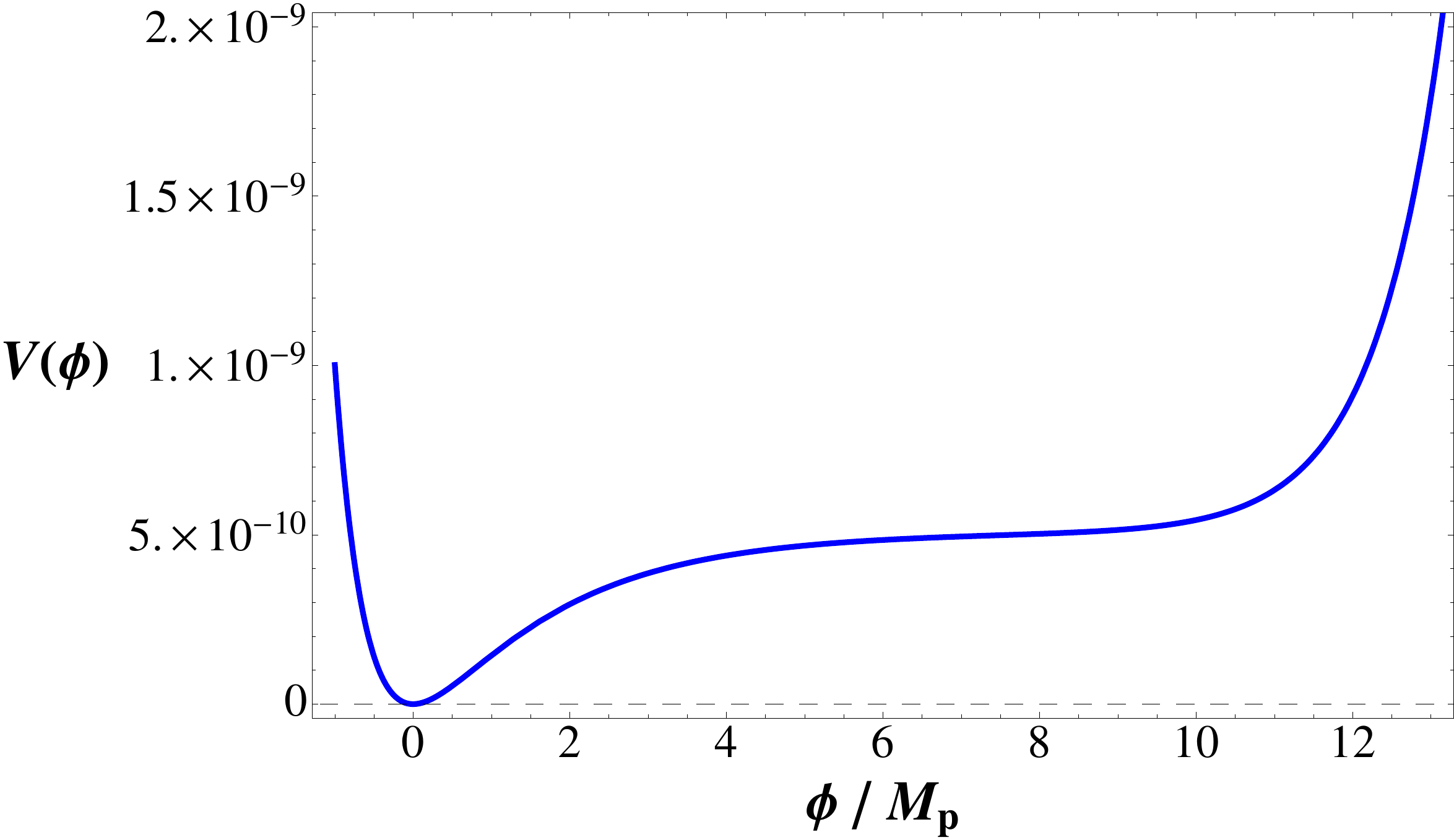}}
 \caption{Sketch of the scalar potential in fibre inflation. $\phi$ denotes the canonically normalized inflaton scalar field in units of $M_{\rm P}$. We clearly see the rapid steepening of the scalar potential beyond the slow-roll plateau around $\phi=(5\ldots 9)\,M_{\rm P}$.}\label{fig:1}
\end{figure}
%

A scalar potential with this shape does not arise typically in the known set of simple low-energy effective single-field inflation models, as these models are engineered to generate slow-roll flatness in the first place. Corrections to a scalar potential driving slow-roll inflation which typically steepen the potential will arise generically due to dimension-six operators from radiative corrections or integrating out massive states, rendering inflation UV sensitive. Therefore we argue that the hints for lack of power on large scales, if interpreted in the context of single field slow roll, can be attributed to high energy effects that are inherently hard to control within the effective field theory.  Hence, candidate fundamental theories of quantum gravity such as string theory are a suitable place for discussing scalar potentials with the above general shape required for generating just 60 e-folds of slow-roll inflation with suppression of CMB power at low-$\ell$ from a prior fast-roll phase.


String theory lives ten space-time dimensions. To make contact with our four dimensional large-scale space-time, we need to compactify the six extra-dimensions on an undetectably small internal manifold. This process of compactification produces a huge set of possible suitable manifolds, each of which is accompanied by set of massless 4d scalar moduli fields describing the allowed deformation modes of each manifold. Major progress in recent years involved the construction of a whole 'landscape'~\cite{Kachru:2003aw,Susskind:2003kw} of type II string flux compactifications (see e.g.~\cite{Grana:2005jc,Douglas:2006es} for a review) with full moduli stabilization and positive vacuum energy necessary to make contact with cosmologically viable 4d space-time descriptions incorporating  the observed late-time acceleration. Using this stage, a considerable body of recent work then realized models of string inflation driven by either a slow-roll flat region of the moduli scalar potential or higher-derivative kinetic terms originating on mobile D-branes. These models of string inflation in type IIA/IIB string theory can be classified by utilizing either D-brane positions, volume or K\"ahler moduli, or axions arising from higher-dimensional gauge-fields in string theory as inflaton scalar fields (see e.g.~\cite{Baumann:2009ni,Cicoli:2011zz,Burgess:2013sla} for very recent reviews).

Here we will look more closely at the type IIB string model of fibre inflation~\cite{Cicoli:2008gp} as a potential candidate realizing our above criterial for low-$\ell$ CMB power suppression from fast-roll. This is due the fact, that fibre inflation by relying on the extended no-scale structure of the underlying Large Volume Scenario moduli stabilization~\cite{Balasubramanian:2005zx,Conlon:2005ki,Cicoli:2007xp} acquires its scalar potential due to the effects of perturbative string quantum corrections which necessarily drive a rather sudden steepening of the potential beyond the slow-roll plateau (see Fig.~\ref{fig:1} for a sketch of the scalar potential from fibre inflation).

The paper is organized as follows. We motivate the study of asymmetric inflection points, of which fibre inflation is one example,  for large scale power suppression in section ~\ref{sec:asymIP}. In section~\ref{sec:fibre} we shortly review fibre inflation in Large Volume type IIB flux compactifications of type IIB string theory. We then analyze in section~\ref{sec:fastroll} the fast-roll regime of steepening of the scalar potential provided by the string loop corrections. We find the steepening to be insufficient to generate a detectable suppression of CMB power at low-$\ell$ during the first e-fold of the observable $\simeq 60$ e-folds of slow-roll inflation. We then proceed to show in section~\ref{sec:powerkill} that increasing the power of the $\tau_1$ (fibre modulus) dependence of the positively contributing string loop correction can lead to the desired effect. This provides motivation for future work searching for string setups providing the required structure of the string loop corrections in fibre inflation. We finally discuss our results in section~\ref{sec:discuss}

\section{Inflection point inflation and large scale power suppression}
\label{sec:asymIP}

In this section we discuss why in order to have power suppression at low-$\ell$ one needs to have asymmetric inflection point.

In inflationary models built around approximately symmetric inflection points of the form
\be
V=V_0\left(1+\lambda_1 \phi+\frac{\lambda_3}{3}\phi^3+...\right)
\ee
the spectral index can be written as a function of the number of efolds $N_e$ and the total number of efolds in the inflationary region $N_{tot}$ as
\be
n_s=1-4\left(\frac{1}{N_e}-\frac{\pi^2}{3}\frac{N_e}{N_{tot}^2}\right),
\ee
to first approximation. In the context of string inflation such potentials can arise from D3 brane inflation.

Current data points us towards a red power spectrum with $n_s-1\sim -0.04$, which assuming the reheating physics is such that one needs $N_e\sim 60$ implies that the plateau around the inflection point must support around $N_{tot}\sim 172$. The upshot of this is that there is too much expansion before the horizon exit of the CMB scales takes place. Therefore, in a generic small-field cubic symmetric inflection point situation, any non-slow roll dynamics that could lead to a power suppression at low multipoles will only affect scales that have long left the horizon and are therefore unobservable in the CMB spectrum.

We therefore argue that since symmetric inflection points give rise to too much expansion, what is required is a asymmetric potential, that gets steep very close to the inflection point. In the space of string inflation models, fibre inflation provides one concrete realisation of inflation in the vicinity of an asymmetric inflection point. In the next section we briefly review this model, before determining if it is suitably asymmetric to suppress power in Section \ref{sec:fastroll}.

\section{Fibre inflation in a nutshell}
\label{sec:fibre}

Fibre inflation is a model of closed string inflation based on geometries of the form
\be
\vo=\lambda_1 t_1 t_2^2+\lambda_2 t_3^3=\alpha(\sqrt{\tau_1}\tau_2 -\gamma\tau_3^{3/2}),
\label{eq:vol}
\ee
where $t_i$ denote 2-cycle volumes, $\tau_i=\partial \vo/\partial t_i$ 4-cycle volumes, $\lambda_i$ the triple intersection numbers, $\alpha=1/(2\sqrt{\lambda_1})$ and $\gamma=\frac 23 \sqrt{\lambda_1/(3\lambda_2)}$. This is one of the few models of string inflation that yields observable primordial gravitational waves. Here we will briefly describe the inflationary potential and phenomenology associated with this model in the context of the large volume scenario (LVS).

By considering $\alpha'^3$ corrections to the K\"ahler potential and non-perturbative effects supported on the cycle $\tau_3$
\be
K=-2\log\left(\vo+\frac{\hat{\xi}}{2}\right)
\qquad\text{and}\qquad
W=W_0+A e^{-a \tau_3},
\label{eq:KandW}
\ee
one can show that a F-term scalar potential gets generated for the K\"ahler moduli:
\be
V_{LVS}=\frac{8 \sqrt{\tau_3} A^2 a^2 e^{-2 a \tau_3}} {\vo}-\frac{4 W_0 \tau_3 A a e^{- a \tau_3}}{\vo^2}+\frac{3 \xi}{4 \,g_s^{3/2}\,\vo^3}\,.
\ee
This potential admits a LARGE volume vacuum  at
\be
\langle\vo\rangle\sim e^{1/g_s} \qquad\text{and}\qquad\langle\tau_3\rangle\sim\frac{1}{g_s}.
\ee

Since $V_{LVS}$ depends on $\tau_1$ only through $\vo$, the physics that stabilises the volume of the compact space leaves a flat direction in the $(\tau_1, \tau_2)$ plane. This apparent setback has the interesting consequence of allowing for the creation of a mass hierarchy in the moduli sector of the compactification, which sets the foundations for an inflationary model in the K\"ahler moduli sector that is effectively a single field. For a more detailed discussion of the mass hierarchy in fibered compactifications see \cite{Cicoli:2008gp,Cicoli:2011ct}.

At the level of eq. (\ref{eq:KandW}) $\tau_1$ is exactly massless, however it is known that $\tau_1$ dependent terms in the $V$ can, and in general will, be generated by perturbative corrections to the K\"ahler potential. From the point of view of string compactifications these corrections arise from the exchange of both open and closed strings between stacks of branes. While an explicit computation of such terms on Calabi-Yau backgrounds is missing, they are conjectured to take the form
\be
\delta K_{g_s}=\delta K_{g_s}^{KK}+\delta K_{g_s}^{W}
\ee
where 
\be
\delta K_{g_s}^{KK}=\sum^{h_{(1,1)}}_{l=1}\frac{C_i^{KK} a_{il} t^l}{Re(S) \vo}
\label{eq:deltaKKK}
\ee
originates from the exchange of closed strings carrying KK momentum between D3 and D7 branes and
 \be
\delta K_{g_s}^W=\sum^{h_{(1,1)}}_{l=1} \frac{C_i^W}{ a_{il} t^l\vo}
\label{eq:deltaKW}
 \ee
arises from exchange of winding strings between D7 branes. Note that in both $C_i^{W}$ and $C_i^{KK}$ are functions of the complex structure moduli. Since complex structure are stabilised at tree level by fluxes, it is reasonable to assume that these subleading terms in their potential will not significantly perturb their vacua.

The scalar potential can be computed as a series expansion in powers of $\delta K$, 
\be
\delta V_{gs}=\sum_i \left(g_s^2{ C_i^{KK}}^2 K^0_{ii}-2 \delta K^W_{g_s, \tau_i}\right)\frac{W_0^2}{\vo^2}.
\ee
In what follows we assume a scenario in which the brane setup (which determines the matrices $a_{ij}$ in eqs. (\ref{eq:deltaKKK}) and (\ref{eq:deltaKW})) is such that the scalar potential is given by
\be
\delta V_{gs}=\left (\frac{(g_s C_1^{KK})^2}{\tau_1^2}-2\frac{C_{12}^W}{\vo \sqrt{\tau_1}}+2\frac{(\alpha g_s C_2^{KK})^2 \tau_1}{\vo^2}\right)\frac{W_0^2}{\vo^2}.
\ee
This potential has the double virtue of stabilising the fibre modulus at 
\be
\frac{1}{\langle\tau_1\rangle^{3/2}}=\frac{4 \alpha C_{12}^W}{(g_s C_1^{KK})^2 \vo}\left(1+sign(C_{12}^W)\sqrt{1+4g_s^4\left(\frac{C_1^{KK} C_2^{KK}}{C_{12}^W}\right)^2}\right)
\ee
and of having a flat plateau suitable for inflation. 

In order to study the inflationary dynamics it is crucial to canonically normalise the inflaton field. Recalling that the complexified K\"ahler moduli are given as $T_i=\tau_i+i b_i$, the kinetic part of the Lagrangian is:
\be
\mathcal{L}_{\rm kin}=K_{i\bar{j}}\partial_{\mu} T_i \partial^{\mu} \bar{T}_j
=\frac{1}{4}\frac{\partial ^2K }{\partial \tau_i \partial \tau_j}
(\partial_{\mu} \tau_i \partial^{\mu} \tau_j+\partial_{\mu} b_i \partial^{\mu} b_j)\,,
\ee

The K\"ahler metric that follows from  Eq. \ref{eq:vol} is then:
\begin{equation}
K^0_{i \bar{j}}=\left(
\begin{array}{ccc}
\frac{1}{4 \tau _1^2} & \frac{\tau _3^{3/2}}{4 \tau _1^{3/2} \tau _2^2} & -\frac{3 \sqrt{\tau _3}}{8 \tau _1^{3/2} \tau _2} \\
 \frac{\tau _3^{3/2}}{4 \tau _1^{3/2} \tau _2^2} & \frac{1}{2 \tau _2^2} & -\frac{3 \sqrt{\tau _3}}{4 \sqrt{\tau _1} \tau _2^2} \\
-\frac{3 \sqrt{\tau _3}}{8 \tau _1^{3/2} \tau _2} & -\frac{3 \sqrt{\tau _3}}{4 \sqrt{\tau _1} \tau _2^2} &\frac{3}{8 \sqrt{\tau _1} \tau _2 \sqrt{\tau _3}}
\end{array}
\right)\,,
\end{equation}
where we kept only the leading order term in the volume expansion for each entry.
The mass hierarchy in the K\"ahler moduli sector explained above allows one to consider the volume and the blow-up cycle as fixed during inflation, simplifying the kinetic term for the fibre modulus to:
\be
\mathcal{L}_{\rm kin}=\frac{3}{8\tau_1^2}\partial_\mu \tau_1 \partial^\mu \tau_1\,.
\ee
It then follows that the canonically normalised field is defined as:
\be
\phi\equiv \frac{\sqrt{3}}{2} \ln \tau_1\qquad \text{or} \qquad \tau_1\equiv e^{\kappa \phi}\qquad \text{with} \qquad \kappa=\frac{2}{\sqrt 3}.
\label{eq:phitau11}
\ee

The scalar potential for the canonically normalised fibre modulus is then
\be
\delta V_{gs}=\frac{W_0^2}{\vo^2} \left ((g_s C_1^{KK})^2 e^{-2 \kappa \phi} -2\frac{C_{12}^W}{\vo}e^{-\frac{1}{2}\kappa \phi}+2\frac{(\alpha g_s C_2^{KK})^2 }{\vo^2} e^{\kappa \phi}\right).
\ee
One final step is required in order to have the right phenomenology: that is to uplift the potential by adding to it a constant contribution of the same magnitude as its depth at the vacuum, after which it becomes
\be
V=V_0\,\left(1-C_{1/2}e^{-\kappa\phi/2}+C_2 e^{-2\kappa\phi}+C_1 e^{\kappa\phi}\right)\quad.
\label{eq:Vinf}
\ee

The inflationary phenomenology of this potential was analysed in  ~\cite{Cicoli:2008gp}, where it was shown to give rise to a sufficiently long period of exponential expansion. The resulting perturbations can be generated at the right scale by adjusting the volume of the compactifications to be $\vo\sim 10^{3}$. The spectral index lies in the observationally preferred region $n_s\sim 0.96$ and the tensor to scalar ratio can be as high as $r\sim0.005$. In this analysis the last term of Eq. (\ref{eq:Vinf}) is assumed to be small enough in the inflationary region as to play no relevant role in it. It is natural to assume this is the case, since its coefficient is of the order of $(g_s/\vo)^2$. Here however we are interested in the possible effects of such a term for the power suppression at large scales, corresponding to the initial stages of the inflationary epoch. We turn to this issue in the next section.

\section{Fast-roll regime in fibre inflation}
\label{sec:fastroll}

Expanding $\phi$ around the minimum, the scalar potential of fibre inflation reads~\cite{Cicoli:2008gp}
\beq
V=V_0\,\left(1-C_{1/2}e^{-\kappa\phi/2}+C_2 e^{-2\kappa\phi}+C_1 e^{\kappa\phi}\right)\quad.
\eeq
Requiring the post-inflationary minimum to sit at $\phi=0$ and $V(\phi=0)=0$ fixes $C_{1/2}=4/3$ and $C_2=1/3$. The slow-roll plateau can be than be adjusted by tuning $C_1\sim g_S^\#,\# >0$ small via choosing $g_S$. Accommodating the observed 60 e-folds at $n_s<1$ requires $C_1\lesssim 10^{-5}$. We see that the potential always contains an asymmetric inflection point at $\phi_{ip}$ determined by two exponentials changing slowest at large $\phi$
\beq
e^{-\kappa\phi_{ip}/2}=3C_1 e^{\kappa\phi_{ip}}\quad.
\eeq
The best hope for steepening beyond the 60 e-fold point $\phi_{60}$ is by having $\phi_{60}=\phi_{ip}$ since for $\phi>\phi_{ip}$ we have $\epsilon$ monotonically increasing.

\medskip

\noindent For $\phi>\phi_{ip}$ consequently the $e^{\kappa\phi}$-term dominates the scalar potential, so we have
\beq
V\simeq V_{ip} \left(1+C_1e^{\kappa\phi}\right)\quad.
\eeq
Computing the slow-roll parameters in this region we find
\beq\label{eq:slowrollpars}
\epsilon=\frac{\kappa^2C_1^2}{2}e^{2\kappa\phi}=\frac38 \eta^2\quad{\rm for}\quad \phi>\phi_{ip}\quad.
\eeq
Hence, we have $n_s>1$, at least for a large range of field values $\phi>\phi_{ip}$. Therefore, for reasons of comparison with CMB data (observations say $n_s(\phi_{60})<1$ at more than $5\sigma$) we cannot put $\phi_{60}>\phi_{ip}$.

We now determine the point $\phi_\delta>\phi_{ip}$ where $\epsilon_\delta>\epsilon_{ip}$ has a value such that $\Delta_{\cal R}^2(\phi_\delta)=\frac{\delta}{100}\Delta_{\cal R}^2(\phi_{ip})$, i.e. the power at $\phi_\delta$ is suppressed to $\delta$ \% of the power at CMB scales. Using eq.~\eqref{eq:slowrollpars} we get that
\beq
e^{-\kappa(\phi_\delta-\phi_{ip})}=\sqrt{\frac{\epsilon_{ip}}{\epsilon_\delta}}=\sqrt{\frac{\Delta_{\cal R}^2(\phi_\delta)}{\Delta_{\cal R}^2(\phi_{ip})}}=\frac{\sqrt\delta}{10}\quad.
\eeq
Hence we see that $C_1 e^{\kappa(\phi_\delta-\phi_{ip})}\ll1$ for typical values $10<\delta<100$, and thus potential stays essentially constant between $\phi_{ip}$ and $\phi_\delta$ with the decrease in power completely driven by the increase in $\epsilon$ above $\phi_{ip}$.

To render this effect visible in the low-$\ell$ region of the CMB power spectrum, the fast rolling phase must take place close to horizon exit. It is therefore crucial not only to ensure that the potential is sufficiently steep over the adequate range in field space but also that this region is immediately followed by the 60 e-foldings region of the potential. One should then estimate the number of e-folds elapsing between the regime of suppressed power at $\phi_\delta$ and CMB scales around $\phi_{ip}$ and require it to be small. This yields
\bea\label{eq:DeltaNe}
\Delta N_e^{(\delta)}&=&\int\limits_{\phi_{ip}}^{\phi_\delta}\frac{{\rm d}\phi}{\sqrt{2\epsilon}}=\frac{1}{\kappa C_1}\int\limits_{\phi_{ip}}^{\phi_\delta}{\rm d}\phi e^{-\kappa\phi}\nonumber\\
&=& \frac{1}{\kappa\sqrt{2\epsilon_{ip}}}\,\left(1-e^{-\kappa(\phi_\delta-\phi_{ip})}\right)\nonumber\\
&=& \frac{1}{\kappa\sqrt{2\epsilon_{ip}}} \left(1-\frac{\sqrt\delta}{10}\right)\quad.
\eea
Using $n_s\gtrsim 0.94$ from the CMB data, limits $\epsilon_{ip}\lesssim 0.01$. Using this conservative value and demanding 
a suppression of large-scale power to $\delta$\%$=50$\%, we get (using that at the inflection point $n_s(\phi_{ip})=1-6\epsilon_{ip}$)
\beq
\Delta N_e^{(50\%)}\gtrsim 3 \sqrt{\frac{0.06}{1-n_s}}\quad.
\eeq
That is, the regime of power suppression is spread out over too many e-folds and the full desired level of power suppression occurs too early to be visible within the earliest single observable e-fold corresponding to $2\leq\ell \lesssim40$. 

We therefore conclude that the positive exponential term in the original fibre inflation example is just not steep enough to provide the necessary rapid power loss of low-$\ell$. The explanation of this apparent feature of the CMB spectrum requires considerably steeper potentials beyond the 60-efolds point.

\section{Suppressing low-$\ell$ power in fibre inflation}
\label{sec:powerkill}

In looking at eq.~\eqref{eq:DeltaNe} we note a way to cure the problem of the preceding section: if we increase the value of $\kappa$ in the $C_1e^{\kappa\phi}$-term in the scalar potential by a factor of a few, then $\Delta N_e^{(\delta)}\sim \kappa^{-1}$ drops accordingly. That is, if we find a setup for fibre inflation with a scalar potential
\beq
V=V_0\,\left(1-C_{1/2}e^{-\kappa\phi/2}-C_2 e^{-2\kappa\phi}+\tilde C_1 e^{\tilde\kappa\phi}\right)\quad.
\eeq
then repeating the derivation of the last section gives us
\bea\label{eq:DeltaNe2}
\Delta N_e^{(\delta)}&=&\int\limits_{\phi_{ip}}^{\phi_\delta}\frac{{\rm d}\phi}{\sqrt{2\epsilon}}=\frac{1}{\tilde\kappa \tilde C_1}\int\limits_{\phi_{ip}}^{\phi_\delta}{\rm d}\phi e^{-\tilde\kappa\phi}\nonumber\\
& =& \frac{1}{\tilde\kappa\sqrt{2\epsilon_{ip}}}\,\left(1-e^{-\tilde\kappa(\phi_\delta-\phi_{ip})}\right)\nonumber\\
&=& \frac{1}{\tilde\kappa\sqrt{2\epsilon_{ip}}} \left(1-\frac{\sqrt\delta}{10}\right)\quad.
\eea
Hence, choosing $\tilde\kappa\gtrsim 3 \kappa=2\sqrt 3$ will generate a sufficient amount of CMB power suppression within the first single observable e-fold corresponding to $2\leq\ell \lesssim40$. Recalling the discussion in Section 2, one sees that the relevant term in the scalar potential arose from a string loop term of the form
\beq
\delta V\sim \frac{\tau_1}{{\cal V}^4}\quad.
\eeq
Hence, the proposed change could arise from a string correction scaling like
\beq
\delta V\sim \frac{\tau_1^{\tilde\kappa/\kappa}}{{\cal V}^p}\quad{\rm with}\quad p>4\;,\; \frac{\tilde\kappa}{\kappa}\gtrsim 3\quad.
\eeq

\begin{figure}[ht!]
	\centering
	\begin{minipage}[b]{0.49\linewidth}
	\centering
\includegraphics[width=1\textwidth]{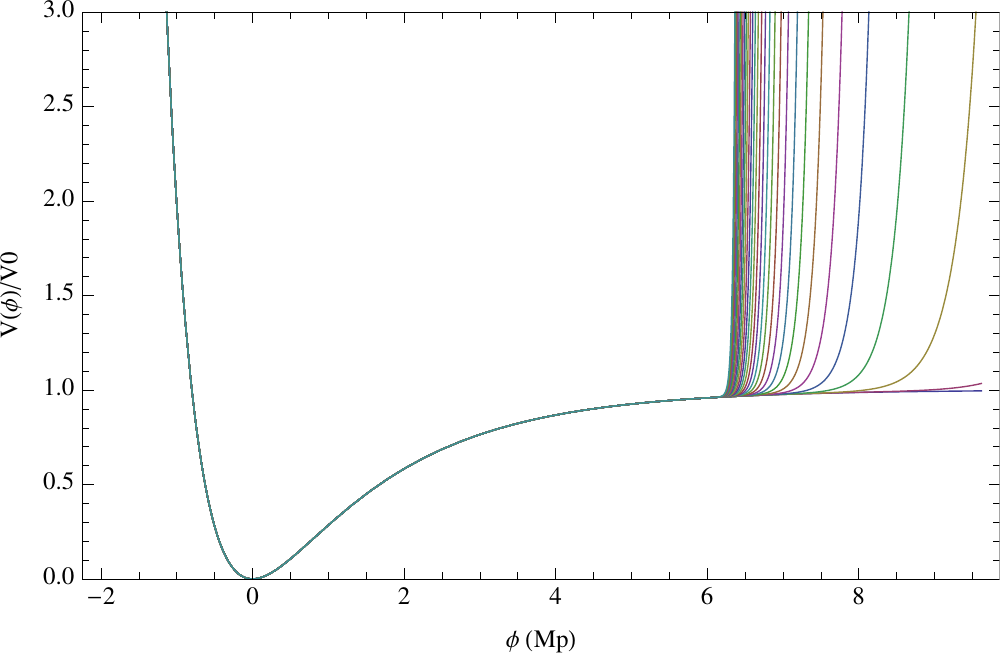}
    \end{minipage}
	\hspace{0.05cm}
	\begin{minipage}[b]{0.49\linewidth}
	\centering
\includegraphics[width=1\textwidth]{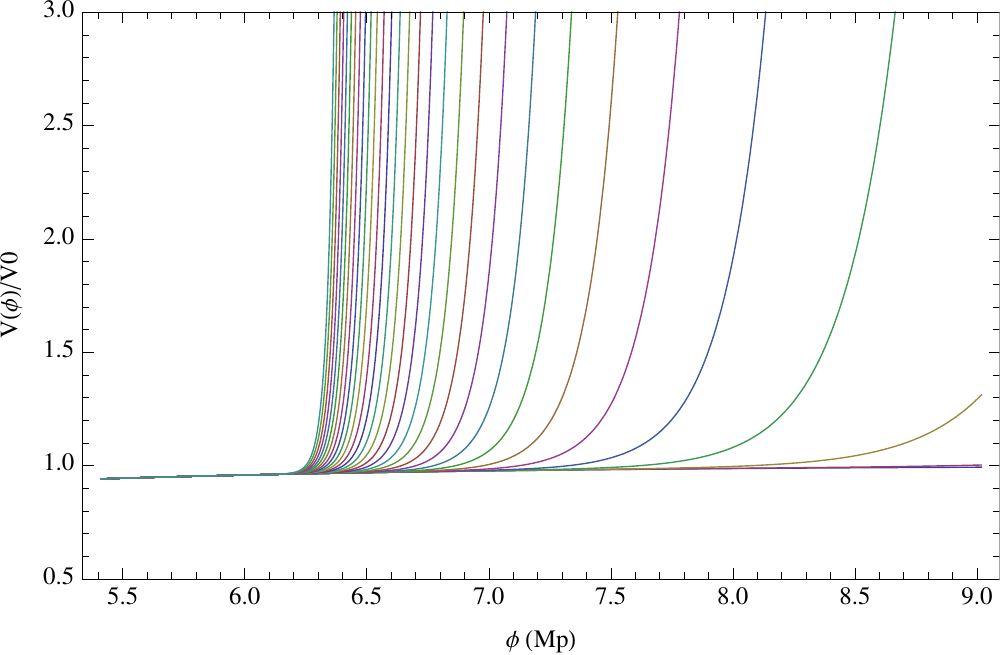}
	\end{minipage}
	\hspace{0.05cm}
	\caption{Fibre modulus potential with different powers in the positive exponential. As $ \frac{\tilde\kappa}{\kappa} $ grows the potential becomes steeper to the right of the inflationary region and the inflection point approaches the 60 e-folding point }
	\label{fig:V_Mod}
\end{figure}

\begin{figure}[ht!]
	\centering
	\begin{minipage}[b]{0.49\linewidth}
	\centering
\includegraphics[width=1\textwidth]{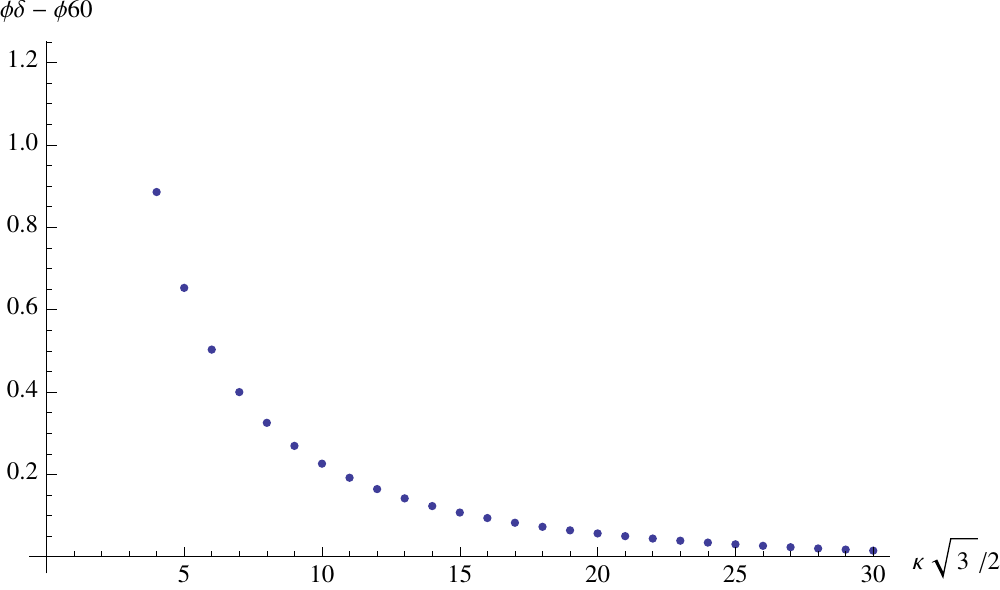}
    \end{minipage}
	\hspace{0.05cm}
	\begin{minipage}[b]{0.49\linewidth}
	\centering
\includegraphics[width=1\textwidth]{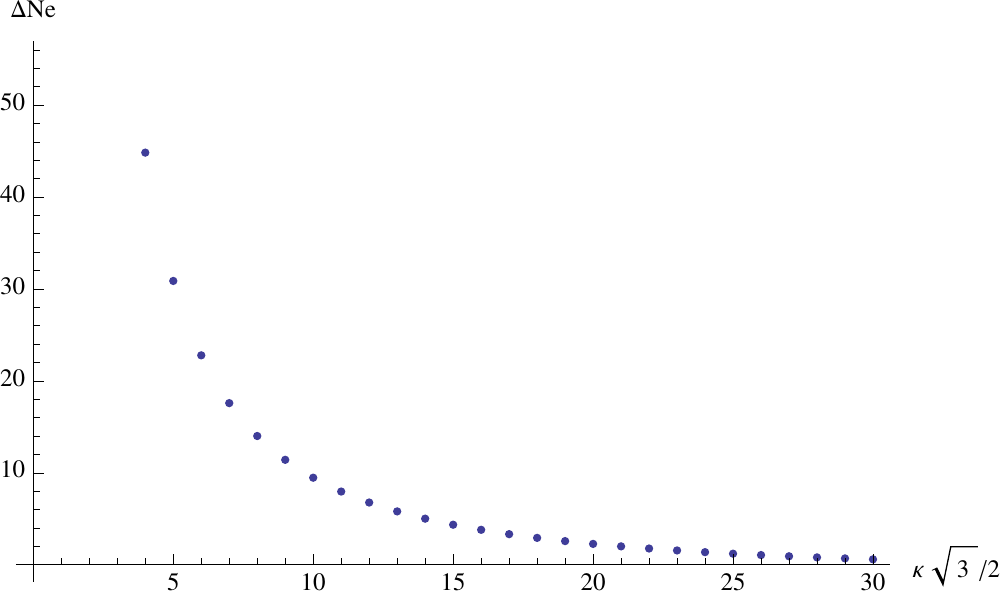}
	\end{minipage}
	\hspace{0.05cm}
	\caption{Left: distance in field space between horizon exit $\phi_{60}$ and the power suppression scale $\phi_\delta$  as a function of the coefficient $\tilde{\kappa}$. For concreteness we assume $\delta\%=50\%$ power suppression. Right: number of e-foldings supported in the region $[\phi_{60},\phi_{\delta}]$ as a function of the coefficient $\tilde{\kappa}$.}
	\label{fig:deltaPhiNe}
\end{figure}

The effect of different choices of $ \frac{\tilde\kappa}{\kappa} $ is illustrated in figure  \ref{fig:V_Mod} for  $ 1\le\frac{\tilde\kappa}{\kappa} \le30$. We have tuned the overall coefficient $C_1$ such that $C_1 \:e^{\tilde{\kappa}\phi}$ would always give a negligible contribution to the potential at horizon exit $\phi=\phi_{60}$.

The effect of steeper exponentials is to move the inflection point towards $\phi_{60}$ therefore reducing the number of e-foldings between the power suppression region $\phi_{\delta}$ and the start of inflation. This is illustrated in figure \ref{fig:deltaPhiNe}.

We can now compare this behaviour using two illustrative numerical examples. The first is the original scalar potential of fibre inflation ($\tilde\kappa=\kappa$) with $C_1\simeq 10^{-5}$. This choice maximizes the slope of the inflationary plateau, pushing $\phi_{N_e}=\phi_{ip}$ while increasing the slope further by lowering the amount of observable slow-roll to $N_e\simeq 50$ by assuming intermediate reheating temperatures.

For the second example we choose $\tilde\kappa=10\kappa$ for the modified loop term, and then set $\tilde C_1=7\times 10^{-33}$ for the same reasons as given above.


%
%

%
\begin{figure}[t!]
 \centerline{\includegraphics[width=0.49\textwidth]{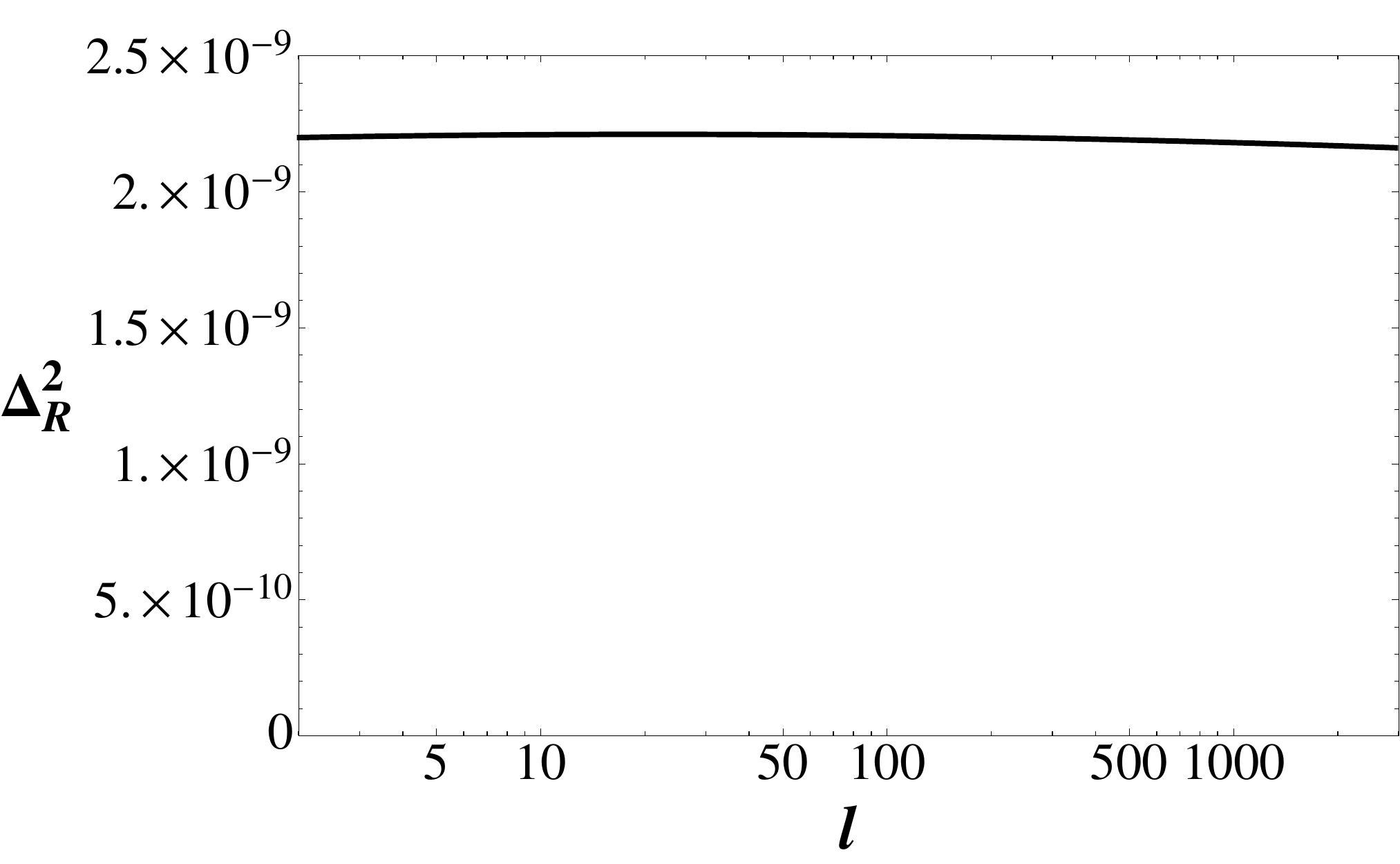}\hspace*{2ex}
\includegraphics[width=0.49\textwidth]{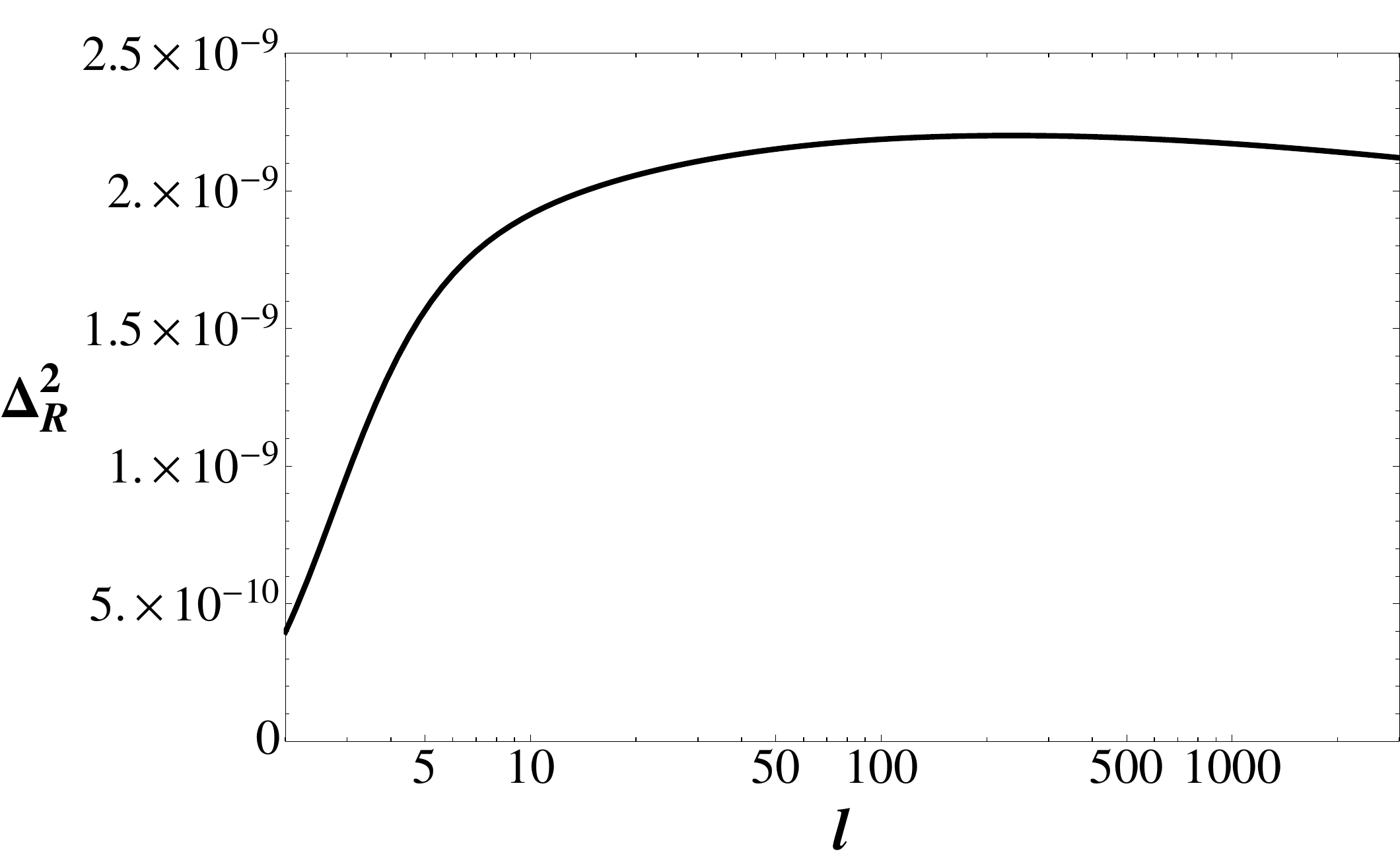}}
 \caption{Power spectrum $\Delta_{\cal R}^2$ computed on the numerical solution $\phi(N_e)$. Left: The original fibre inflation setup with $\tilde\kappa=\kappa=2/\sqrt 3$ and $C_1=10^{-5}$. Right: The modified setup with $\tilde\kappa=10\kappa$ and $\tilde C_1=7\times 10^{-33}$. The extra steepening leads to a clear suppression of curvature perturbation power at low $\ell$.}\label{fig:3}
\end{figure}
%

We can then solve the scalar field equation of motion
\beq
\phi''+3\,\left(1-\frac16 \phi'^2\right)\left(\phi'+\frac{1}{V}\frac{\partial V}{\partial \phi}\right)=0\quad{\rm with}\quad()'\equiv \frac{d}{dN_e}()
\eeq
numerically and compute the slow-roll parameters
\beq
\epsilon=\frac12 \left(\frac{\partial V/\partial\phi}{V}\right)^2\quad,\quad\eta=\frac{\partial^2 V/\partial\phi^2}{V}
\eeq
on the numerical solution $\phi(N_e)$.
This allows us to identify the point $\phi_\delta>\phi_{ip}$ where $\epsilon_\delta=\frac{100}{\delta}\epsilon_{ip}$ suppressing the power there to $\Delta_{\cal R}^2(\phi_\delta)=\frac{\delta}{100}\Delta_{\cal R}^2(\phi_{ip})$. For the above choice of $\tilde\kappa$ and $\tilde C_1$ we can show that a suppression level of $\delta = 50\%$ still allows for $\eta_{\delta}\lesssim 0.1$. This keeps the the moderately faster-rolling region immediately prior to the flat plateau within the slow-roll approximation, and allows us to compute the curvature perturbation in slow-roll. Fig.~\ref{fig:3} shows the power spectrum
\beq
\Delta_{\cal R}^2=\frac{1}{4\pi^2}\frac{H^4}{\dot\phi^2}
\eeq
of the curvature perturbation evaluated on the numerical solution $\phi(N_e)$ within the region corresponding the CMB scales $2<\ell<3000$ for the two examples. We clearly see that fibre inflation per se does not generate appreciable suppression of power in the first $2-3$ observable e-folds of inflation, while the modified version with $\tilde\kappa=10\kappa$ can generate a sizable and sufficiently rapidly varying lack of power at $\ell < 40$.

\section{Discussion}
\label{sec:discuss}

Features in the 2-point function at large scales open a window to the earliest stages of the inflationary epoch. If observed with enough statistical significance, these features can become an extra constraint on inflationary model building, to be added to the usual discriminants: the spectral index and the tensor-to-scalar ratio. 

In this paper we have analysed one such feature that might have been observed in the CMB power spectrum: the lack of power at low-$\ell$. We have argued that it can be generated by a phase of moderately fast roll immediately preceding the usual slow roll inflationary phase. We have shown that an asymmetric inflection point is required to model this behaviour and performed a detailed analysis of one such model: fibre inflation. 

Fibre inflation's string loop generated potential, even though asymmetric, is not steep enough to successfully suppress power at low-$\ell$. The reason being that though a power suppression region exists, it is located too far from the slow roll dynamics region, and so the inflationary expansion corresponding to inflaton motion between these two regions will render any initial power suppression unobservable in the CMB.

In an attempt to understand the structure of the potential required to accommodate this feature, we deviate from string derived fibre inflation and introduce a steeper exponential in the potential. By tuning it to be negligible in the inflationary plateau and to quickly dominate for larger $\phi$, we show that such potential can give rise to the desired level of power suppression.

\noindent {\bf Note added:} This paper is submitted simultaneously to the related work~\cite{Michele1}.

\section*{Acknowledgments}
We are grateful to Michele Cicoli for discussions and to Giulia Costoloni for her patience.
This work was supported by the Impuls und Vernetzungsfond of the Helmholtz Association of German Research Centres under grant HZ-NG-603, and German Science Foundation (DFG) within the 
Collaborative Research Center 676 "Particles, Strings and the Early Universe".





\bibliographystyle{JHEP.bst}
\bibliography{low-l}
\end{document}